\documentclass[iop]{emulateapj}

\newcommand{\xmm} {{\it XMM-Newton}}
\newcommand{\chandra} {{\it Chandra}}
\newcommand{\nustar} {{\it NuSTAR}}
\newcommand{\swift} {{\it Swift}}
\newcommand{\erosita} {{\it eROSITA}}
\newcommand{\swiftxrt} {{\it Swift}/XRT}

\newcommand{\cmsq} {cm$^{-2}$}
\newcommand{\nh} {$N_{\rm{H}}$}
\newcommand{\lx} {$L_{\rm{X}}$}

\newcommand{\chisq} {$\chi^2$}

\newcommand{\rchisq} {$\chi^2_r$}

\newcommand{\degree}{{$^\circ$}}
\newcommand{\ergs}{\mbox{\thinspace erg\thinspace s$^{-1}$}}
\newcommand{\ergcms}{\mbox{\thinspace erg\thinspace cm$^{-2}$\thinspace s$^{-1}$}}
\newcommand{\cntrt}{counts\,s$^{-1}$}

\newcommand{\newulx} {M51 XT-1}

\newcommand{\nobs}{105}

\shorttitle{Swift monitoring of M51}
\shortauthors{Brightman et al.}

\begin{document}

\title{Swift monitoring of M51: A 38-day super-orbital period for the pulsar ULX7 and a new transient ULX}

\author{Murray Brightman$^{1}$, Hannah Earnshaw$^{1}$, Felix F\"{u}rst$^{2}$, Fiona A. Harrison$^{1}$, Marianne Heida$^{3}$, Gianluca Israel$^{4}$, Sean Pike$^{1}$, Daniel Stern$^{5}$, Dominic J. Walton$^{6}$}

\affil{$^{1}$Cahill Center for Astrophysics, California Institute of Technology, 1216 East California Boulevard, Pasadena, CA 91125, USA\\
$^{2}$European Space Astronomy Centre (ESA/ESAC), Operations Department, Villanueva de la Ca\~{n}ada (Madrid), Spain\\
$^{3}$European Southern Observatory, Karl-Schwarzschild-Stra{\ss}e 2, 85748 Garching bei M\"{u}nchen, Germany\\
$^{4}$INAF--Osservatorio Astronomico di Roma, via Frascati 33, 00078 Monteporzio Catone, Italy\\
$^{5}$Jet Propulsion Laboratory, California Institute of Technology, Pasadena, CA 91109, USA\\
$^{6}$Institute of Astronomy, Madingley Road, Cambridge CB3 OHA, UK\\}

\begin{abstract}

We present the results from a monitoring campaign made with the {\it Neil Gehrels Swift Observatory} of the M51 galaxies, which contain several variable ultraluminous X-ray sources (ULXs). The ongoing campaign started in May 2018, and we report here on $\sim1.5$ years of observations. The campaign, which consists of \nobs\ observations, has a typical cadence of 3--6 days, and has the goal of determining the long-term X-ray variability of the ULXs. Two of the most variable sources were ULX7 and ULX8, both of which are known to be powered by neutron stars that are exceeding their isotropic Eddington luminosities by factors of up to 100. This is further evidence that neutron star powered ULXs are the most variable. Our two main results are, first, that ULX7 exhibits a periodic flux modulation with a period of 38 days varying over a magnitude and a half in flux from peak to trough. Since the orbital period of the system is known to be 2 days, the modulation is super-orbital, which is a near-ubiquitous property of ULX pulsars. Secondly we identify a new transient ULX, \newulx, the onset of which occurred during our campaign, reaching a peak luminosity of $\sim10^{40}$ \ergs, before gradually fading over the next $\sim200$ days until it slipped below the detection limit of our observations. Combined with the high-quality \swiftxrt\ lightcurve of the transient, serendipitous observations made with \chandra\ and \xmm\ provide insights into the onset and evolution of a likely super-Eddington event.

\end{abstract}


\section{Introduction}

The galaxies of M51, which consist of the main spiral galaxy M51a/NGC5194 and its smaller companion M51b/NGC5195, are a host to a multitude of X-ray sources \citep[e.g.][]{brightman18b}. These include two active galactic nuclei (AGN) powered by accreting supermassive black holes (SMBHs), one at the center of each of the galaxies; several ultraluminous X-ray sources (ULXs); and transient sources, such as the type IIb supernova SN~2011dh.

Both SMBHs are weakly accreting, relatively speaking, with Eddington ratios $<10^{-4}$, and the AGN powered by the SMBH in M51a is heavily obscured by Compton thick (CT) material. The ULXs, in contrast, are believed to be strongly accreting compact objects. Indeed, two of them are known to be powered by neutron stars (NSs) far less massive than the SMBHs but apparently equally as powerful. For ULX8, the NS identification was determined from the detection of a cyclotron resonance scattering feature (CRSF), caused by interactions of charged particles with a strong magnetic field \citep{brightman18}. For ULX7, this was determined from the detection of coherent pulsations from the source \citep{rcastillo19}. 

The observational properties of ULXs originally suggested they were good intermediate-mass black hole candidates \citep[e.g.][]{earnshaw16}. However, the detection of coherent pulsations and CRSFs mean that some of them certainly are not black holes since black holes are unable to produce such signals. ULXs powered by NSs are a surprising phenomenon since their luminosities imply extreme Eddington ratios of up to 500, and are not a well understood phenomenon. 

The first of NS ULX to be identified was M82~X-2 \citep{bachetti14}, followed by NGC~7793~P13, NGC~5907~ULX1 and NGC~300~ULX1 \citep{fuerst16,israel17a,israel17,carpano18}, with the most recent discovery being NGC~1313~X-2 \citep{sathyaprakash19}. Thus M51 ULX7 and ULX8 are among a small but growing group of these sources. Being hosted in the same galaxy and separated by a few arcminutes means they can be easily studied together.

No pulsations have been detected from ULX8 yet, and hence a spin period is unknown. The spin period of ULX7 is $\sim2$\,s, which is similar to the $\sim1$\,s spin periods for M82~X-2, NGC~7793~P13 and NGC~5901~ULX1. These sources also show another distinguishing characteristic: X-ray flux modulations with periods of 60--80 days. These were all detected from long term monitoring by {\it The Neil Gehrels Swift Observatory} \citep[hereafter \swift,][]{walton16,hu17,qiu15,fuerst18}. Theories to explain these invoke a precession of the accretion disk or wind and some geometric beaming \citep[e.g.][]{dauser17,middleton18}. 

Furthermore, NGC~7793~P13 and NGC~5901~ULX1 exhibit unusual `off' states in addition to the periodic modulations, where the X-ray flux drops significantly, by orders of magnitude, relative to their peak brightness \citep{motch14,walton15}. While the nature of these off states is currently unknown, \cite{tsygankov16} suggest they are related to the onset of the propeller effect, in which the magnetospheric radius moves outside the co-rotation radius such that accretion is dramatically suppressed and the X-ray flux drops precipitously. 

Since the discoveries that M51 harbors two ULXs powered by neutron stars, we have obtained monitoring of the galaxies with \swift\ in order to identify potential periodic flux modulations and off states exhibited by the other sources of this class. We present the results of this monitoring campaign here. In addition to ULX7 and ULX8, ULX4 was identified as a source of interest since it exhibits a bimodal distribution of fluxes, possibly related to the propellor effect \citep{earnshaw18}, and \cite{urquhart16} identified two eclipsing ULXs in the galaxy. Finally, M51 has also hosted several transient phenomena such as the type II supernova SN~2005cs, the type IIb supernova SN~2011dh and the intermediate luminosity red transient AT~2019abn \citep{jencson19}. The \swift\ monitoring also allows us to study these other interesting ULXs and to look for transients in the X-ray band. We assume a distance of 8.58 Mpc to M51 throughout this study \citep{mcquinn16}.  During the preparation of this manuscript, \cite{vasilopoulos19b} was published presenting results similar to ours regarding ULX7 based on the publicly available \swift\ data. Our analyses differ, however, since they use the online tool described in \cite{evans09}, to calculate the XRT count rates, whereas we use our own tailored analysis method. We also present $\sim100$ days more data than they do, covering an extra three cycles of the modulation.

\section{Swift/XRT data analysis}
\label{sec_swift}

We have conducted a systematic monitoring campaign of the M51 galaxies with \swiftxrt\ \citep{burrows05} since 2018-05-01 with a typical cadence of 3--6\,days and typical exposure time of 2000\,s per snapshot. Since our main goal was to investigate the long-term X-ray lightcurves of the two NS-powered ULXs, ULX7 and ULX8, to look for periodic flux modulations, the observing campaign was designed with this goal in mind. However, as described above, the M51 galaxies contain many other interesting X-ray sources which we have also obtained long-term lightcurves for.

\begin{figure*}
\begin{center}
\includegraphics[width=180mm]{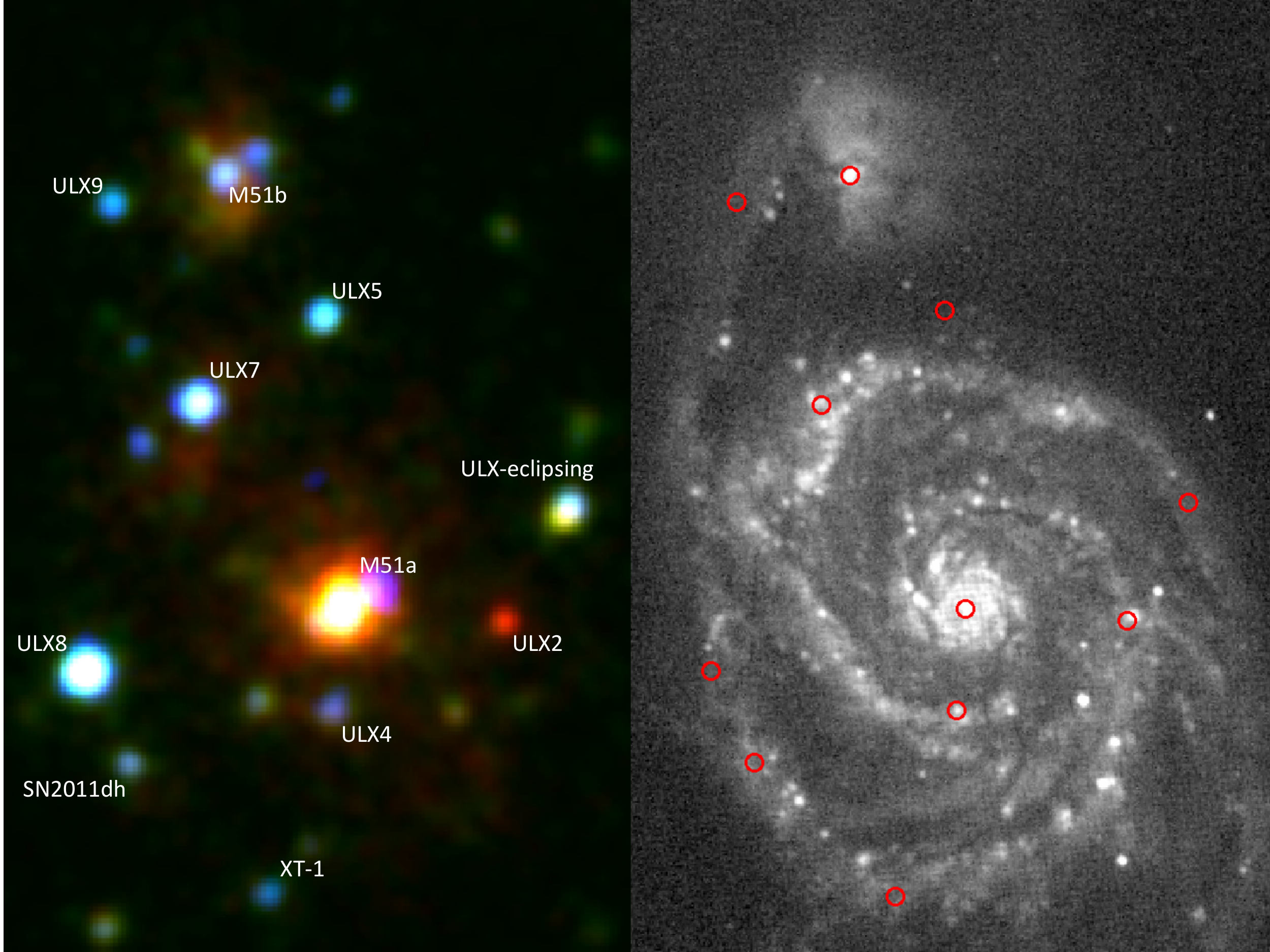}
\caption{Left - Stacked \swiftxrt\ images of the M51 galaxies. Red represents 0.3--1 keV emission, green represents 1--2.5 keV emission and blue represents 2.5--10 keV emission. The image has been smoothed with a 4\arcsec\ Gaussian and the sources of study have been labelled. Right {\it Swift}/UVOT obsID 00032017112 {\it u}-band image of M51 with the same sources marked with red circles }
\label{fig_xrt_img}
\end{center}
\end{figure*}

In order to determine which additional sources are bright enough to provide useful lightcurves, we stack all \nobs\ observations using the online XRT analysis tool. We create images in three bands, 0.3--1 keV, 1--2.5 keV, and 2.5--10 keV. The image is presented in Figure \ref{fig_xrt_img} where red, green and blue represent these three bands respectively. The AGN of the two galaxies, M51a and M51b, several known ULXs are clearly detected, as well as SN~2011dh and a previously uncatalogued ULX which we call \newulx.

\begin{figure}
\begin{center}
\includegraphics[width=90mm]{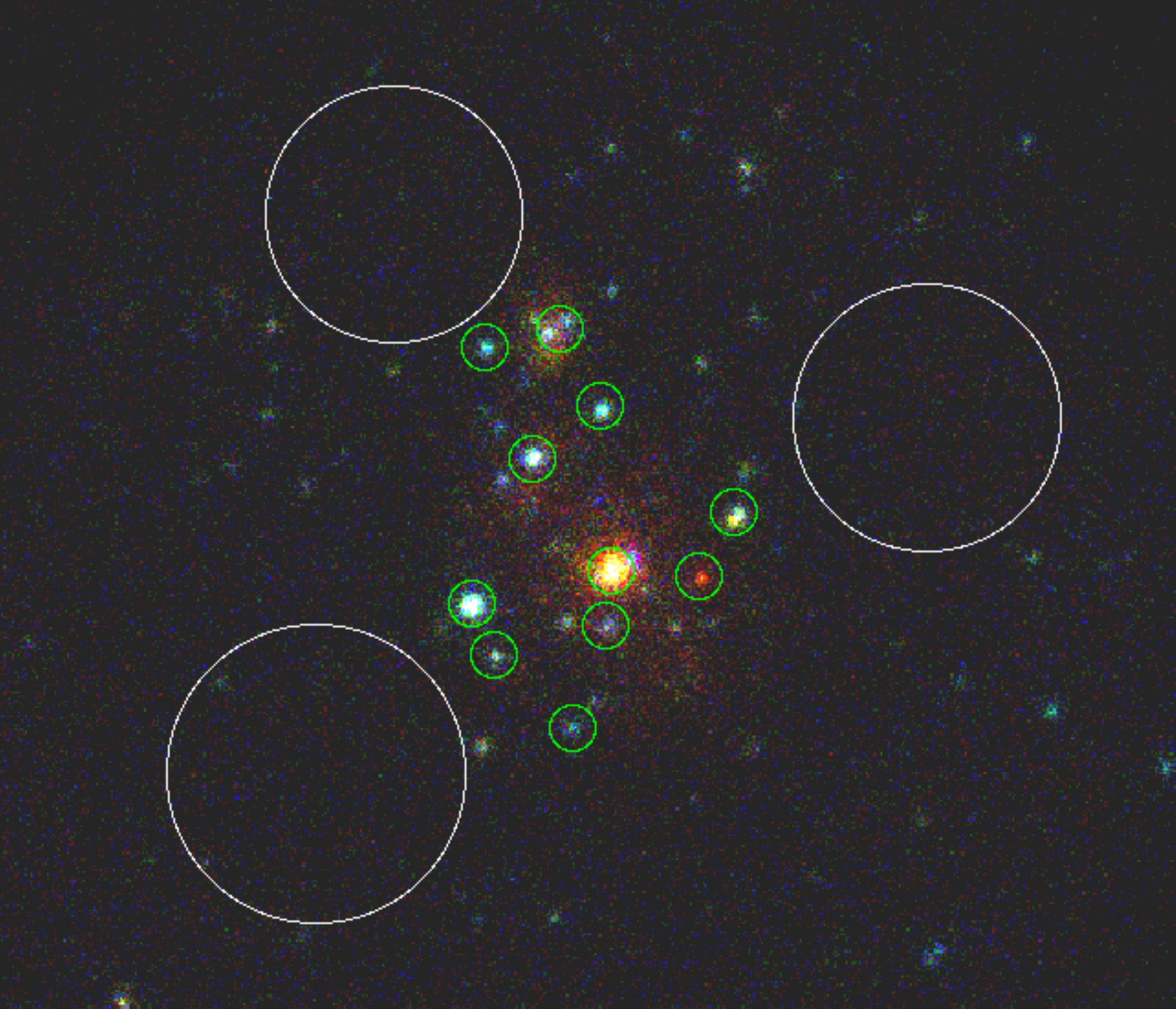}
\caption{Same as Figure \ref{fig_xrt_img} (left), but image is wider angle, not smoothed and shows the source (green) and background (white) extraction regions.}
\label{fig_xrt_img2}
\end{center}
\end{figure}

We proceed to extract spectra from all sources using the {\sc heasoft} v6.25 tool {\sc xselect}. Source events are selected from a circular region with a 25\arcsec\ radius and background regions consisting of large circles external to the galaxies are used to extract background events. All extraction regions are shown in Figure \ref{fig_xrt_img2}. For each source spectrum, we construct the auxiliary response file (ARF) using {\tt xrtmkarf}. The relevant response matrix file (RMF) from the CALDB is used. All spectra are grouped with a minimum of 1 count per bin.

\begin{figure*}
\begin{center}
\includegraphics[trim=10 40 10 10, width=180mm]{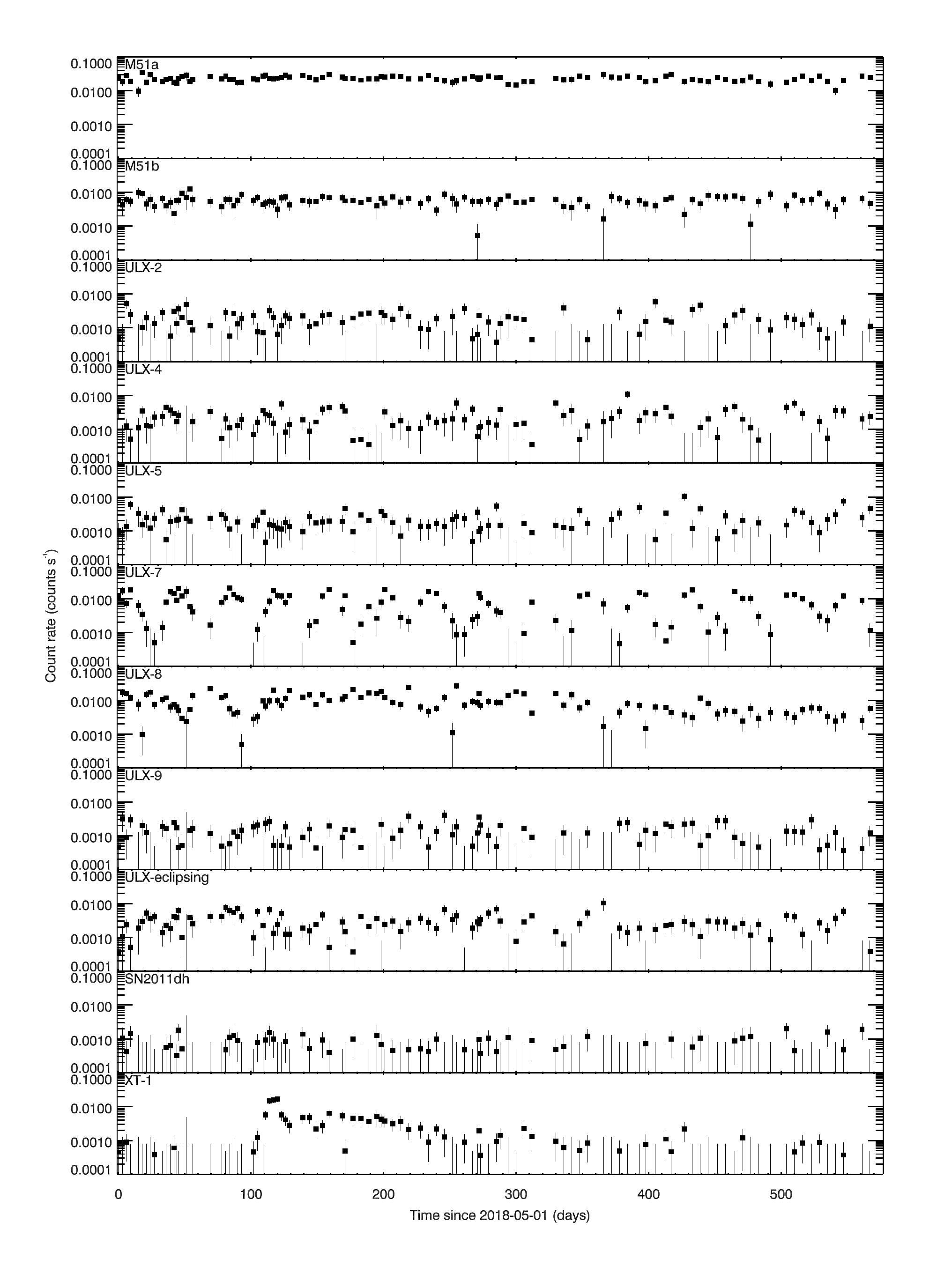}
\caption{0.3--10 keV \swiftxrt\ lightcurves of all sources of interest in M51 presented in \cntrt\ as black squares with 1$\sigma$ error bars. 90\% upper limits for non-detections are shown as error bars without squares.}
\label{fig_xrt_ltcrv}
\end{center}
\end{figure*}

We use the {\sc heasoft} tool {\sc xspec} to calculate background-subtracted count rates in the 0.3--10 keV band. The lightcurves of these sources are shown in Figure \ref{fig_xrt_ltcrv}. For observations where a source has zero total counts, we estimate the 90\% upper limit on the count rate using a typical background count rate of 7$\times10^{-5}$ \cntrt\ and Poisson statistics.

Our next step is to look for count rate variability in the sources. We do this by testing for deviations from a constant count rate by calculating the \chisq\ of the count rates where $$\chi^2\equiv\sum_{n=1}^{N_{\rm obs}}\left(\frac{CR_{n}-\langle CR\rangle}{\sigma_n}\right)^2.$$ $CR_n$ is the count rate in each observation, $n$, $\langle CR\rangle$ is the mean count rate averaged over all observations, and $\sigma_n$ is the 1$\sigma$ uncertainty on the count rate for each observation. We present the results in Table \ref{table_stats}. We find that the sources which exhibit evidence for count rate variability, which we arbitrarily define as \rchisq$\equiv$\chisq$/N_{\rm obs}>2.0$, are ULX4, ULX7, ULX8, the eclipsing ULXs, and \newulx. The sources with the largest variability (i.e. \rchisq$>10.0$) are indeed ULX7 and ULX8.

\begin{table*}
\centering
\caption{Source statistics}
\label{table_stats}
\begin{center}
\begin{tabular}{l l l c r c r}
\hline
Source name	& Position & SIMBAD indentifier & $\langle CR\rangle$	& \chisq	& $N_{\rm obs}$ & \rchisq \\
(1) & (2) & (3) & (4) & (5) & (6) & (7)\\
\hline
M51a&13:29:52.773,+47:11:42.08&M51&0.0226& 148.5&         105& 1.414\\
M51b&13:29:59.221,+47:15:58.57&NGC 5195&0.0057& 176.5&         105& 1.681\\
ULX-2&13:29:43.449,+47:11:35.23&RX J132943+47115&0.0015& 163.7&         105& 1.559\\
ULX-4&13:29:53.300,+47:10:42.00&[TW2004] NGC 5194 37&0.0021& 262.1&         105& 2.496\\
ULX-5&13:29:53.937,+47:14:38.40&RX J132954+47145&0.0020& 201.2&         105& 1.916\\
ULX-7&13:30:01.098,+47:13:42.33&RX J133001+47137&0.0073&2770.0&         105&26.381\\
ULX-8&13:30:07.458,+47:11:05.62&RX J133007+47110&0.0088&1117.9&         105&10.647\\
ULX-9&13:30:06.015,+47:15:42.27&RX J133006+47156&0.0012& 130.9&         105& 1.246\\
ULX-eclipsing&13:29:39.861,+47:12:44.75&RX J132939+47126&0.0027& 364.6&         105& 3.472\\
SN2011dh&13:30:04.993,+47:10:11.21&SN 2011dh&0.0004&  45.3&         105& 0.431\\
XT-1&13:29:56.843,+47:08:52.60&-&0.0015& 395.6&         105& 3.768\\

\hline
\end{tabular}
\tablecomments{Summary of the sources studied here. Column (1) lists the source name adopted, column (2) gives the position in J2000 co-ordinates, column (3) lists the SIMBAD identifier, column (4) gives the mean 0.3--10 keV \swiftxrt\ count rate in \cntrt, column (5) shows the \chisq\ of the count rate, column (6) gives the number of observations used, and column (7) gives the reduced \chisq.}
\end{center}
\end{table*}

Following this, we look for periodic flux modulations from all the sources with evidence for variability. We use two techniques to do so. First we do epoch folding, whereby the lightcurve is folded on several test periods and deviations from a constant flux are determined \citep{leahy97}. This technique is well suited for finding periodic signals which are not necessarily sinusoidal. We also use a Lomb-Scargle periodogram which is commonly used to search for periodicities in unevenly sampled data and is specialized for finding sinusoidal signals \citep{lomb76,scargle82}. 

\begin{figure}
\begin{center}
\includegraphics[trim=10 10 10 10, width=80mm]{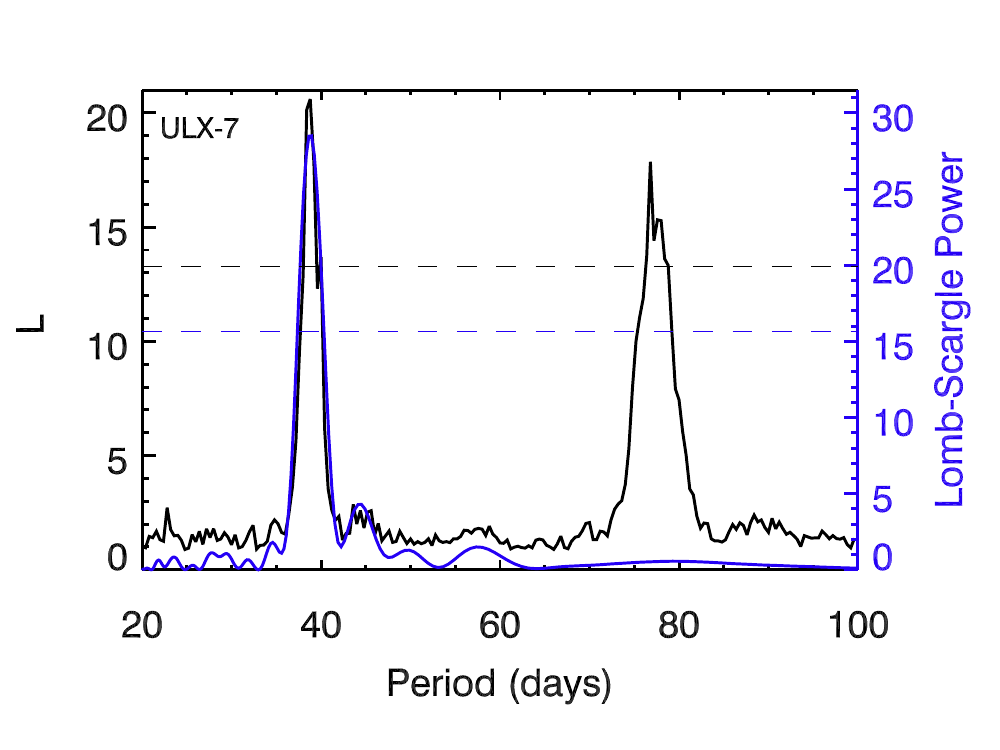}
\caption{Results from the epoch-folding (black) and Lomb-Scargle analysis (blue) on the lightcurve of ULX7. A clear signal at 38 days is seen in both periodograms of this source, which is significant at $>99$\% significance, based on simulations (dashed lines).}
\label{fig_epfold}
\end{center}
\end{figure}

For the epoch folding analysis, we search for periods ranging from 20--100 days over 200 linearly spaced steps. For each test period we split the lightcurve into 10 phase bins, as has been common practice in these searches and compute the L-statistic \citep{davies90}. For the Lomb-Scargle analysis, we search over the same period range with the same number of test periods. As shown in Figure \ref{fig_epfold}, ULX7 has a strong peak at 38-days in the periodogram seen from both epoch folding and the Lomb-Scargle analysis. A peak at twice this period is also seen in the epoch folding periodogram, which is likely a harmonic of the shorter period peak, but we discuss the possibility that it is the fundamental later. There are no strong peaks seen in any other of the variable sources.

\subsection{Significance simulations}

In order to determine the significance of the signals exhibited in the periodograms, we conduct simulations to determine the false alarm rate, i.e. the rate at which such a signal could be produced from a statistical fluctuation rather from a real signal. While both the epoch folding using the L-stat and the Lomb-Scargle periodogram define tests to assess the significance of signals, they can often overestimate the significance. Therefore, we follow \cite{walton16b} by simulating 10,000 lightcurves with 2000s resolution and a red noise power spectrum and we sample them with the same observational strategy as the real lightcurves, running the same analysis as described above. We then note the largest peak in each periodogram, irrespective of period. We determine the 99.9\% significance level by calculating what level 99.9\% of the simulated periodograms fall under. We plot this quantify on Figure \ref{fig_epfold}, which shows that the 38-day period is detected with a significance greater than 99.9\%.

\section{Investigating the periodic flux modulation from ULX7}

The orbital period of the NS powering ULX7 and its binary companion is known to be 2 days, which was determined from the timing analysis of the pulsations \citep{rcastillo19}. Therefore the 38-day periodic flux modulation we detect here from ULX7 is super-orbital in nature, making it the fourth ULX pulsar where this characteristic has been identified. Interestingly, ULX8 does not exhibit any periodic flux modulations on the timescales we have searched.

\begin{figure*}
\begin{center}
\includegraphics[trim=50 10 10 10, width=190mm]{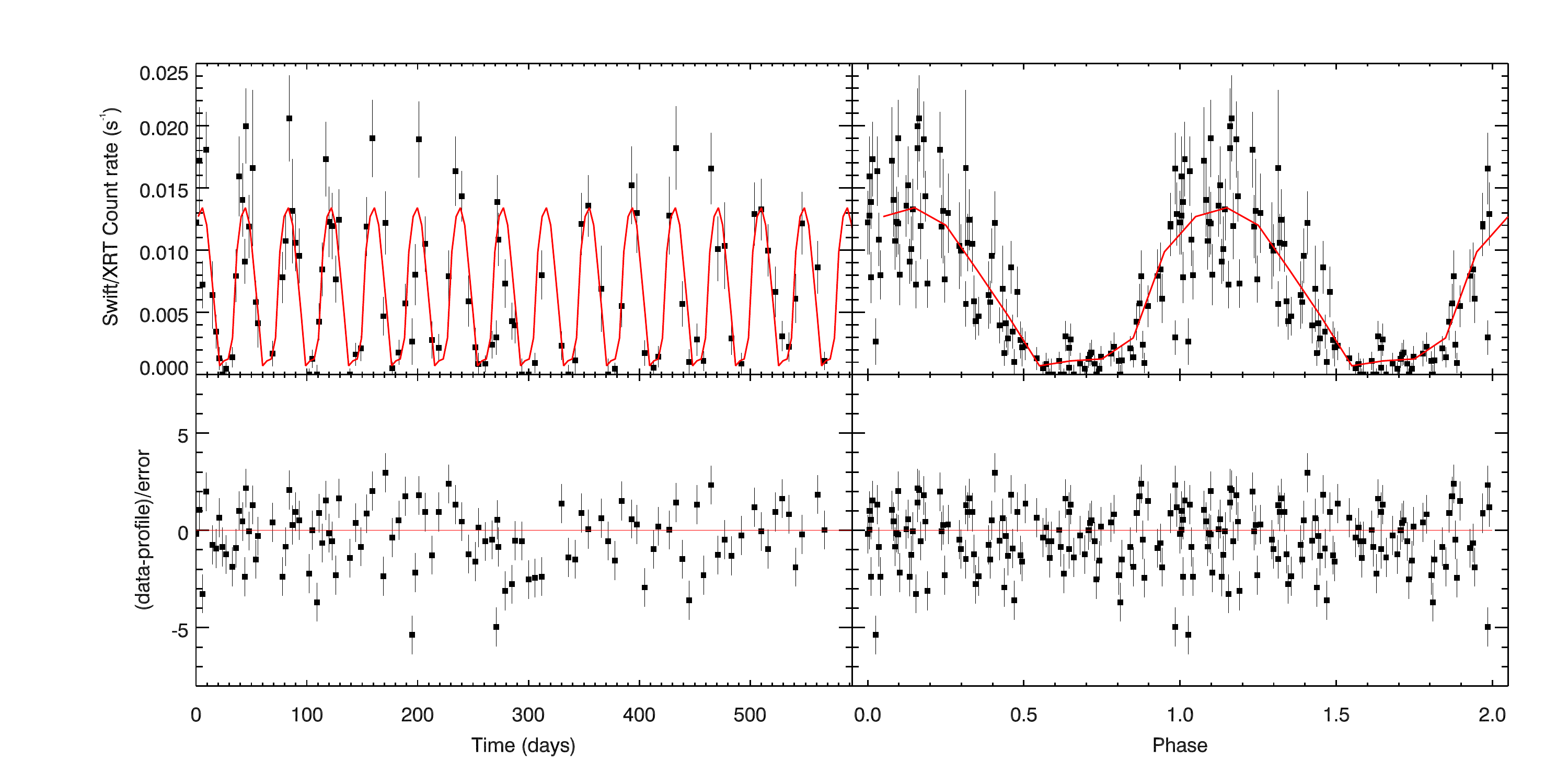}
\caption{Top left - 0.3--10 keV \swiftxrt\ lightcurve of ULX7 (black data points, 1$\sigma$ errors), overplotted with the average 38-day profile determined from epoch folding (red line). Top right - The same lightcurve but folded on the 38-day period and plotted against phase. Bottom left - Deviations of the data from the average profile shown against time. Bottom right - deviations of the data from the average profile folded on the 38-day period.}
\label{fig_ulx7_epfold}
\end{center}
\end{figure*}

Investigating further, from the epoch folding analysis described above, we determine the average profile of the modulation by calculating the mean count rate in each of the 10 phase bins. The average profile is plotted in Figure \ref{fig_ulx7_epfold} and appears sinusoidal. The mean profile of the modulations peaks at $\sim0.013$ \cntrt\ and has a minimum at $\sim0.001$ \cntrt. This is more than an order of magnitude from peak to trough, and corresponds to a luminosity range of $\sim8\times10^{38}$--$10^{40}$ \ergs.

The deviations from this average profile, which we define as $\sigma=\frac{data-profile}{error}$ are also shown. These do not appear to show any long term structure in addition to the sinusoidal modulation, with the exception of three dips in the lightcurve where $\sigma<-4$. Also plotted in Figure \ref{fig_ulx7_epfold} are the average profile and data plotted against phase, as well as the deviation from the profile. The dip features are clustered at phase$\sim0.8$--1, but further monitoring is required to determine if these features are coherent in phase of the flux modulation.

\begin{figure}
\begin{center}
\includegraphics[trim=50 10 10 10, width=80mm]{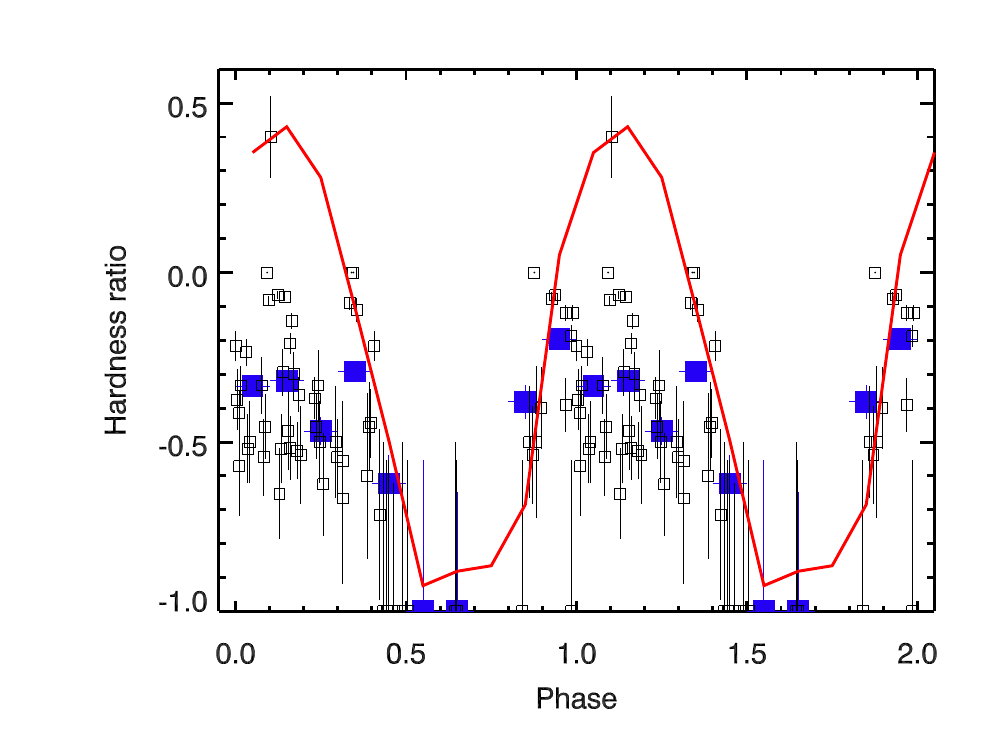}
\caption{Hardness ratios of ULX7 as a function of phase of the super-orbital period (black squares), with binned averages (blue open squares). The scaled average profile of the count rate is shown for reference (red line).}
\label{fig_hr}
\end{center}
\end{figure}

We next look for spectral variations as a function of the phase of the flux modulation, which may yield insight into their origin. While the individual $\sim2$\,ks \swiftxrt\ observations do not have the photon statistics to conduct spectral analysis, we calculate hardness ratios for each, where $HR=(H-S)/(H+S)$, $S$ is the total number of counts in the 0.3--2 keV band and $H$ is the total number of counts in the 2--10 keV band. We show the results in Figure \ref{fig_hr}, where we also display the binned averages. We find that the source is relatively hard during the peaks of the modulation, and relatively soft during the troughs.

\section{A new transient ULX}

In addition to the well-known bright X-ray sources in M51, we detect a new, previously uncatalogued source when stacking the XRT images (Figure \ref{fig_xrt_img}). The new source appeared in the southern part of M51a at coordinates 13h 29m 56.8s, +47\degree\ 08\arcmin\ 52.6\arcsec\ (J2000), and we refer to it as \newulx. The lightcurve of the source in Figure \ref{fig_xrt_ltcrv} shows that it was a transient. The source was weakly detected, or undetected during the first $\sim100$ days of the monitoring campaign, with a count rate $<0.001$ \cntrt, but became one of the brightest X-ray sources in the galaxy with a count rate of $\sim0.02$ \cntrt\ in the space of 10--15 days. The peak of its activity occurred during a 6-day period covering 2018-08-23 and 2018-08-29. The source then gradually faded over the next $\sim200$ days and was once again weakly detected at day $\sim$400 of our monitoring campaign.

There are no known, catalogued sources at the position of \newulx, with no sources within 20\arcsec\ in the \chandra\ source catalog of \cite{kuntz16}, which is the deepest catalogue of X-ray sources in M51 available, summing over 800 ks of \chandra\ exposure time. There are also no matching sources in the latest ULX catalogue of \cite{earnshaw19} compiled from the \xmm\ serendipitous source catalog nor in the \xmm\ serendipitous source catalog itself (3XMM DR8).

We searched the Transient Name Server\footnote{https://wis-tns.weizmann.ac.il} for known transients in M51, finding none that occurred in 2018. AT2019abn was the closest in time and spatial separation and was an intermediate luminosity red transient discovered on 2019-01-22 several arcmins from \newulx\ \citep{jencson19}. Figure \ref{fig_xrt_img} shows the \swift/UVOT $u$-band image of M51 during the peak of the outburst: no source is seen at the position of \newulx. We also looked at the UVOT data from later observations to check for delayed emission at longer wavelengths, but we did not see anything either.

\begin{figure}
\begin{center}
\includegraphics[trim=10 10 10 10, width=80mm]{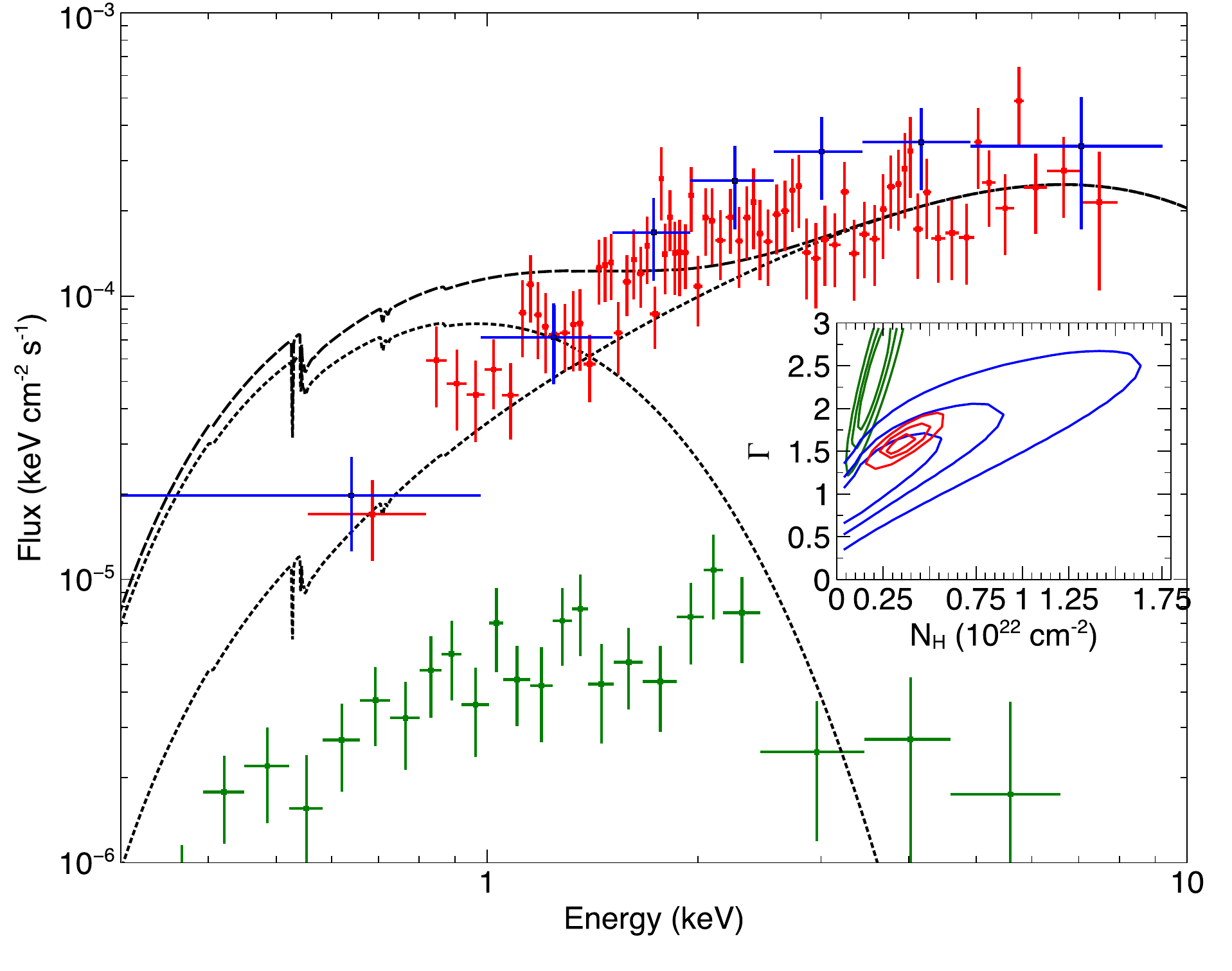}
\caption{\swiftxrt\ unfolded spectrum of \newulx\ during the peak of its outburst (blue), the \chandra\ unfolded spectrum from several days afterwards (red), and the \xmm\ unfolded spectrum from $\sim1$ year later (green). The dashed black lines show the spectrum of ULX7 for comparison, from a joint \xmm\ and \nustar\ observation, which consists of two disk-like components, each shown with dotted lines, as commonly observed in ULXs. This shows that the spectrum of \newulx\ at its peak resembles the hotter disk-like component seen in ULXs, with the cooler component missing, while the spectrum has become considerably softer at late times. Inset are confidence contours resulting from a fit with an absorbed power-law to the \swift, \chandra, and \xmm\ spectra of \newulx.}
\label{fig_j1329_spec}
\end{center}
\end{figure}

Using the build \swiftxrt\ spectral products tool\footnote{http://www.swift.ac.uk/user\_objects/} \citep{evans09}, we created a spectrum of \newulx\ when it was at the peak of its outburst. We grouped the spectrum with a minimum of 1 count per bin and fitted it with an absorbed powerlaw. The spectrum is shown in Figure \ref{fig_j1329_spec}. The spectrum is very hard with $\Gamma=1.1^{+0.7}_{-0.5}$ and no evidence for absorption above the Galactic value. The source had a peak flux of 1.2$\times10^{-12}$ \ergcms, which corresponds to a luminosity of 1.1$\times10^{40}$ \ergs\ at 8.58 Mpc. 

Serendipitously, \chandra\ observed M51 for 19.8 ks on 2018-08-31 (obsID 20998), only a few days after the peak of activity from \newulx. This allowed us to obtain a better position of the source than \swiftxrt\ provided, and a higher signal-to-noise spectrum. We ran the {\sc ciao} tool {\tt wavdetect} on the observation to obtain list of positions for all sources. We then cross correlated this source list with {\it Gaia} DR2 catalog to obtain the astrometric shifts. Having applied the astrometric shifts to the \chandra\ source catalog, we determine the position of \newulx\ to be R.A. = 13:29:56.97, decl. = +47:08:54.1 (J2000), with an astrometric uncertainty of 0\farcs45 from the residual offsets with the {\it Gaia} catalog.

We used the {\sc ciao} v4.11 tool {\sc specextract} to extract the spectrum of the source from a circular region with a 2\arcsec\ radius. Background events were expected from a nearby region. The source had a count rate of 4.1$\pm0.1\times10^{-2}$ \cntrt\ on the ACIS detector, and the spectrum is well fitted by an absorbed power-law spectrum ({\tt tbabs*powerlaw}) where \nh$=3.3\pm0.1\times10^{21}$ \cmsq\ and $\Gamma=1.61\pm0.20$ with a flux of $7.2\pm0.7\times10^{-13}$ \ergcms, which corresponds to a luminosity of 6.3$\times10^{39}$ \ergs\ at 8.58 Mpc. The spectrum is shown in Figure \ref{fig_j1329_spec}.

Additionally, \xmm\ observed M51 on 2019-07-11 (obsID 0852030101), almost a year after the outburst, also serendipitously. The source was detected with a count rate of 4$\times10^{-3}$ \cntrt\ in the pn detector. We extracted a spectrum of the source using a circular region with a radius of 10\arcsec, and extracted the background from a nearby region, using the {\sc xmmsas} v18.0.0 tool {\sc evselect}. The source is described by a power-law spectrum with \nh=1.9$^{+1.1}_{-0.9}\times10^{21}$ \cmsq\ and $\Gamma=2.6\pm0.5$, which presents a clear softening of the spectrum since the outburst. The source exhibited a flux of 2.0$^{+0.4}_{-0.5}\times10^{-14}$ \ergcms, which corresponds to a luminosity of 2$\times10^{38}$ \ergs\ at 8.68 Mpc. The spectrum is also shown in Figure \ref{fig_j1329_spec}.

To place these spectra in context, we show in Figure \ref{fig_j1329_spec} the spectrum of ULX7 which was obtained from a joint \xmm\ and \nustar\ observing campaign in 2019 (Brightman et al. {\it in prep}). The spectrum of ULX7 is similar to that of other ULXs with good quality broadband spectra \citep[e.g.][]{walton17}, consisting of two disk-like components with different temperatures. ULX7 also has a similar flux and luminosity as \newulx\ at its peak, therefore serving as a good comparison. As seen, the spectra of \newulx\ at its peak are harder than ULX7 and resemble the hot disk-like component seen in that source. Although the absorption is higher for \newulx\ (3.3 compared to 0.7$\times10^{21}$ \cmsq), this does not account for all the spectral hardness as a cooler disk-like component can be ruled out in the \chandra\ spectrum. We discuss the physical implications of this further in Section \ref{sec_disc}.

\begin{figure}
\begin{center}
\includegraphics[trim=50 10 50 10, width=80mm]{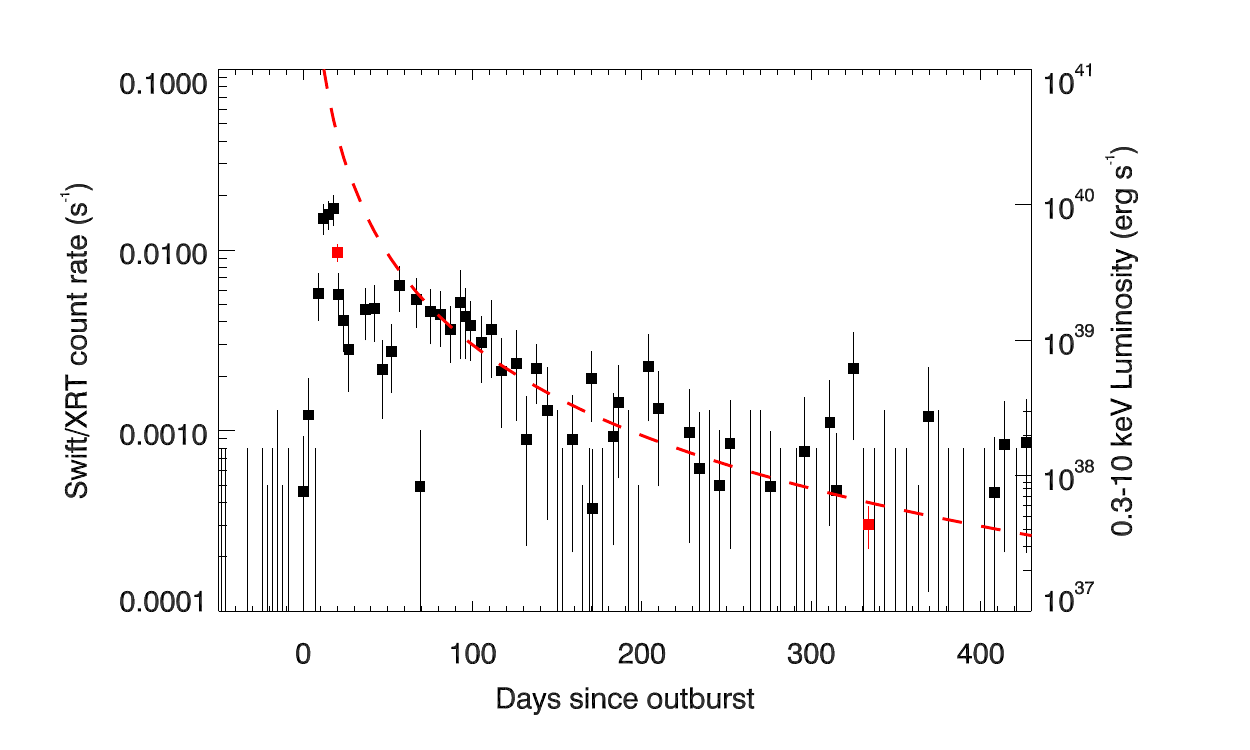}
\caption{Lightcurve of \newulx\ during its outburst monitored by \swiftxrt\ (black squares), and serendipitously observed with \chandra\ and \xmm\ (red squares). Also shown is a red dashed line with a $t^{-5/3}$ shape, which matches the data well after $\sim50$ days of outburst.}
\label{fig_j1329_ltcrv}
\end{center}
\end{figure}

Finally, \xmm\ observed M51 four times (obsIDs 0824450901, 0830191401, 0830191501, and 0830191601) in a period of $\sim100$ days before the outburst, allowing us to place tight upper limits on the flux of the source before outburst. For each individual observation, the upper limit on the 0.3--10 keV flux is $\sim8\times10^{-15}$ \ergcms, corresponding to \lx$\sim7\times10^{37}$ \ergs. Figure \ref{fig_j1329_ltcrv} shows the lightcurve of \newulx\ with the \chandra\ and \xmm\ data added. We also show a line with a $t^{-5/3}$ shape that matches the data well after $\sim50$ days of outburst for discussion in Section \ref{sec_disc}.

Using the astrometrically corrected \chandra\ position of \newulx, we searched the {\it Hubble Source Catalog (HSCv3)} for potential counterparts. Only one source is found within the 0\farcs45 positional uncertainty, at a position of 13:29:56.983, +47:08:54.35 detected by WFC3 in the F110W filter with a magnitude of 24.44. The distance modulus corresponding to 8.58 Mpc is 29.7, implying an absolute magnitude of this source of $M_{\rm F110W}=-5.22$.  

\section{Discussion}
\label{sec_disc}

\subsection{A 38-day super-orbital flux modulation from ULX7}

The orbital period of the NS powering ULX7 and its binary companion is known to be 2 days, which was determined from the timing analysis of the pulsations \citep{rcastillo19}. Therefore the 38-day periodic flux modulation we detect here from ULX7 is super-orbital in nature, making it the fourth ULX pulsar where this characteristic has been identified. Indeed it appears as if this characteristic is near-ubiquitous in these systems. This leads to the possibility that identifying periodic flux modulations in ULXs where the accretor is unknown, can be used as strong evidence that it is powered by a NS. During the preparation of this manuscript, \cite{vasilopoulos19b} also presented the identification of the 38-day super-orbital period of M51 ULX7 from the same \swift\ data we have presented here. 

We list the spin period, orbital period, and super-orbital period for all ULX pulsars where these super-orbital periods have been identified so far in Table \ref{table_sorbitalp} for reference and comparison. M51 ULX7 has the shortest super-orbital period of the four, at 38 days, where the others are at 60--80 days. As mentioned earlier, the epoch folding periodogram presents a peak at twice the period of the 38-day peak, which is most likely a harmonic, but if it is the true period, it would make it comparable to the other ULX pulsars. ULX7 also has the longest spin period  for ULX pulsars with super-orbital periods, at 2.8\,s, whereas the others have 0.4--1.4\,s. A larger sample is needed before any conclusions can be drawn regarding any inverse relationship between the spin period of the NS and the super-orbital period.

The luminosity range of the flux modulation from ULX7, which is $\sim8\times10^{38}$--$10^{40}$ \ergs\ is larger than NGC~7793~P13 and NGC~5901~ULX1 but smaller to M82~X-2. However, due to the observational challenges related to M82~X-2, ULX7 is much easier to study, making it the best source to study these extreme flux variations.

Not only does it appear that these super-orbital periods are a near-ubiquitous property of ULX pulsars, but they have been observed in many other neutron-star X-ray binary systems, such as Her~X-1 \citep{tananbaum72,katz73}, LMC~X-4 \citep{lang81} and SMC X-1 \citep{gruber84} as well as black hole binaries \citep[e.g. compilation by][]{sood07}. In these lower luminosity and lower mass accretion rate systems, a warped precessing disk is often invoked to explain the flux modulations. Here the limb of the warped disk periodically occults the X-ray emitting region, causing the periodic dips in flux.

From our hardness ratio analysis, we found that the source is hard during the peaks of the modulation, and soft during the troughs. This argues against changes in the absorption causing the flux modulation, as this would have predicted the source being hard during the troughs, rather than soft, since absorption causes attenuation of soft X-rays. Occultation of the source by a warped disk would appear to be disfavored, unless the source is completely obscured during the occultations and we only see scattered emission.

Furthermore, it is likely that the thin accretion disk that is needed for warps does not exist in ULX pulsars. In these systems the extreme accretion rate causes the disk to extend vertically giving it a large scale height. Indeed the precession of a large scale height disk has been proposed as the mechanism behind the flux modulations in ULX pulsars \citep[e.g.][]{fuerst17,dauser17,middleton18}. This theory would also fit the observations of M51 ULX7 better. Here, where the source is at low fluxes and soft, we are potentially viewing the disk more edge on where only the cooler outer regions can be seen. When the source is at high fluxes, we see into the hot funnel region. This would require a very large precession angle, however, to go from a disk seen almost face on, to edge on.

\cite{middleton18} suggest that the precession can be explained by the Lense-Thirring effect, a consequence of General Relativity whereby a massive spinning object induces precession of orbiting particles that are displaced vertically from the rotation axis, for example, a large scale height accretion flow. Indeed, \cite{middleton19} suggest that these timescale periodicities can be used to determine whether the compact object is a black hole or a NS.

Four \xmm\ observations of M51 took place during the period 2018-05-13 to 2018-06-15, which covered the first cycle of the flux modulation we observed with \swift. Indeed, it was from these observations that the pulsations confirming ULX7 to be powered by a neutron star were detected \citep{rcastillo19}. \cite{rcastillo19} also presented a spectral analysis of the source from these data, finding too that the source became very soft during the minimum of the flux modulation. They also suggest that the luminosity swing may be caused by a change in inclination angle of the disk, though the scaling of the temperature of the disk with luminosity did not fit this hypothesis. Alternatively they proposed that the periodic flux modulations may be induced by changes in accretion rate caused by the orbit of a third star, although this explanation did not account for the spectral variations.

Interestingly, \cite{earnshaw16} presented the results from a detailed systematic spectral analysis of ULX7 using \xmm, \chandra\ and \nustar, finding no evidence for significant spectral evolution, even at the lowest fluxes. They did however find evidence for soft diffuse thermal emission surrounding the source in the \chandra\ data, that is likely contaminating the larger PSFs of \xmm\ and \swift. It is possible that this component dominates the \xmm\ and \swift\ data at lower fluxes, making the source appear soft.

\begin{table*}
\centering
\caption{Super-orbital periods of all ULX pulsars known to date.}
\label{table_sorbitalp}
\begin{center}
\begin{tabular}{l c c c c c c c c c}
\hline
Source name & $P_{\rm spin}$ & Ref  & $P_{\rm orbit}$ & Ref & $P_{\rm super-orbital}$ & Ref  & \lx,min -- \lx,max & Ref \\
(1) & (2) & (3) & (4) & (5) & (6) & (7) & (8) & (9)\\
\hline
M82 X-2 & 1.37	& 1,2 & 2.5 & 1,2 & 64 & 3 & $\sim10^{38}$--$10^{40}$ & 4,3  \\
NGC~7793~P13 & 0.42	& 5,6 & 63.9 & 7 & 66.8 & 7 & $\sim4\times10^{39}$--$10^{40}$ & 8,7 \\
NGC~5907~ULX1 & 1.43 & 9 & 5.3 & 9 & 78.1 & 10 & $\sim3\times10^{40}$--$10^{41}$ & 10,11\\
M51~ULX7 & 2.8	& 12 & 2.0 & 12 & 38 & 13 & $\sim8\times10^{38}$--$10^{40}$ & 13 \\
\hline
\end{tabular}
\tablecomments{Summary of the observed properties of the ULX pulsars which exhibit super-orbital periods in their lightcurves. Column (1) lists the source name, column (2) lists the spin period of the neutron star in s as determined from the pulsations, column (3) gives the reference for the spin, column (4) gives the orbital period in days of the neutron star and its binary companion determined from timing analyses of the pulsations, column (5) gives the reference for the orbital period, column (6) gives the super-orbital period in days, column (7) gives the reference for this. Column (8) gives the luminosity range of the super-orbital flux modulations in \ergs\ and column (9) gives the reference. References - 1.\cite{bachetti14}, 2. \cite{bachetti19}, 3. \cite{brightman19}, 4. \cite{brightman16}, 5. \cite{fuerst16}, 6. \cite{israel17}, 7. \cite{fuerst18}, 8. \cite{walton17}, 9. \cite{israel17a}, 10. \cite{walton16b}, 11 \cite{fuerst17}, 12. \cite{rcastillo19}, 13. This work.}
\end{center}
\end{table*}

\subsection{The nature of the new transient ULX \newulx}

The peak luminosity and timescale of \newulx, the new transient we have identified, corresponds to the ``ULX in outburst'' section of the X-ray transients luminosity-timescale plot from \cite{soderberg09}. However, we note that the decline of the outburst after $\sim50$ days is consistent with a $t^{-5/3}$ relationship, which is the signature of the fallback from a tidal disruption event (TDE, Figure \ref{fig_j1329_ltcrv}). TDEs are however much brighter and occur in the nuclei of galaxies, thought to be due to the tidal disruption of a star by a supermassive black hole.

Other examples of transient ULXs are NGC~300~ULX1, which had an X-ray outburst in 2010, reaching an X-ray luminosity of 5$\times10^{38}$ \ergs \citep{binder11}. The source was observed at lower fluxes in observations made in 2014 \citep{binder16} but then reached ULX luminosities during observations made in 2016 with \lx$\sim5\times10^{39}$ \ergs\ during which pulsations were detected \citep{carpano18}. Regular \swift\ monitoring of the source in 2018 revealed that the source initially persisted at these luminosities but then went into decline. Spectral analysis showed a hard spectrum.

Another source, Swift~J0243.6+6124, was an X-ray transient found in our own Galaxy, identified by \swift/BAT \citep{cenko17} and with no previously reported activity. The source reached an X-ray luminosity of $2\times10^{39}$ \ergs\ in a period of around 30 days, before steadily declining over a period of $\sim100$ days \citep{wilsonhodge18}. The detection of pulsations identified it as a neutron star accretor \citep{kennea17}. The source exhibited re-brightenings in the X-ray after the decline, albeit to peak luminosities around 2 orders of magnitude less than the initial outburst \citep{vandeneijnden19}.

Most recently, \cite{earnshaw19} identified a new ULX in the galaxy NGC 6946 that was previously undetected in several archival \xmm\ and \chandra\ observations, but caught at a luminosity of $2\times10^{39}$ \ergs\ during a simultaneous \xmm\ and \nustar\ observation in 2017. The source was then undetected 10 days later in a \chandra\ observation. Again the spectrum was hard with $\Gamma\sim$1. Several hypotheses were put forward by \cite{earnshaw19} regarding the nature of this transient event, which may also explain \newulx. These included a supernova, a neutron-star-powered ULX briefly leaving the propeller regime, a black hole binary exhibiting a hard-only outburst, and a micro-tidal disruption event.

A supernova is unlikely to be the cause of \newulx\ since no supernova candidates were announced at its time and position and the galaxy is regularly monitored by transient surveys such as ZTF. Considering the peak luminosity of the outburst, a hard-only outburst would require a black hole much more massive than the black hole binaries in our Galaxy. The similarity of the outburst to other transient neutron star ULXs described above appears to favour that scenario, but the more exotic hypothesis of a micro-tidal disruption event cannot be ruled out, especially considering the $t^{-5/3}$ decline in count rate. It could be possible that the source is in the background of M51 and is infact a nuclear TDE seen at larger distances, and hence luminosities. However, \newulx\ is located in the spiral arm of M51, so the source should be absorbed, which is inconsistent with the X-ray spectrum.

Other examples of transient ULXs are ones in M31 \citep{middleton12,middleton13}, M83 \citep{soria12}, NGC~1365 \citep{soria07}, NGC~3628 \citep{strickland01}, NGC~5907 \citep{pintore18}, NGC~7090 \citep{liu19}, among others. For most of these transient ULXs, only a handful of observations were made and the lightcurves were not well sampled. Here, we serendipitously obtained a high quality lightcurve that captured the onset of activity and covered it until it became undetected. 

Since most ULXs are now considered to be super-Eddington accretors, we have likely witnessed the onset of a super-Eddington event with \newulx\ in M51, and the details we have gleaned from it may give clues regarding the formation of such an accretion flow. Most persistent ULXs show a characteristic broadband spectrum consisting of two disk-like components. \cite{walton17} presented a compilation of ULXs with good quality broadband X-ray spectra, finding that they are remarkably similar, even comparing those with known neutron star accretors and those with unknown accretors. The spectral components consist of two disk-like components, one potentially associated with the hot, large-scale-height inner flow, and the other a cooler component, perhaps associated with the outer part of the disk or the photosphere of an outflow. Furthermore, a high-energy tail is often detected, possibly associated with the accretion column of a pulsar, or Comptonized emission.

When comparing the spectral shape of \newulx\ during the peak of its activity to the ULX population, we noted that it resembled the hot disk-like component attributed to the large-scale-height inner flow of other ULXs, with the cooler component absent (Figure \ref{fig_j1329_spec}). This perhaps suggests that the cooler component had yet to form. If the cooler component is indeed associated with the outer disk, its absence may explain why the outburst was relatively short, since it was not supplied by more material in the outer part of the disk. If instead the cooler component is from an outflow, it would imply that there is a time lag between the formation of the super-Eddington flow and the ejection of the material. Indeed a year after the outburst, the spectrum appeared much cooler, which may be the remnants of the outflow, albeit with the hotter component gone. Following an event such as this in future with more detail will allow a clearer picture of the formation and evolution of a super-Eddington flow. With the launch of \erosita\ \citep{merloni12}, which will observe each part of the sky every 6 months, more of these transient ULXs will undoubtedly be identified.

\section{Summary and Conclusions}
\label{sec_conc}

We have presented the results from the systematic monitoring of M51 by \swiftxrt\ over the period of $\sim1.5$ years which has yielded high-quality X-ray lightcurves of more than 10 X-ray sources, including two AGN and several ULXs. Our main results are the detection of a 38-day super-orbital flux modulation from the ULX pulsar, ULX7, and the identification of a new, transient ULX, \newulx. 

The super-orbital period is a near-ubiquitous property of ULX pulsars, possibly driven by the precession of a large scale height disk. However, the magnitude-and-a-half swings in flux are hard to reconcile with this theory. Alternatively, the spectral shape as a function of phase does not favor periodic obscuration of the source, unless it is being completely obscured and only scattered light is seen.

The outburst of \newulx\ appears similar to super-Eddington outbursts from other neutron stars, suggesting we have witnessed the onset and decline of a highly super-Eddington event. The spectral shape at the peak of the outburst is similar to the hot-disk like component seen in other ULXs that is attributed to a large-scale-height accretion flow.

\section{Acknowledgements}

We wish to thank the \swift\ PI, Brad Cenko for approving the target of opportunity requests we made to observe M51, as well as the rest of the \swift\ team for carrying them out. We also acknowledge the use of public data from the \swift\ data archive. This work made use of data supplied by the UK Swift Science Data Centre at the University of Leicester. The work of DS was carried out at the Jet Propulsion Laboratory, California Institute of Technology, under a contract with NASA. DJW acknowledges support from an STFC Ernest Rutherford Fellowship.

\end{document}